\documentclass[preprint]{aastex631}
\usepackage{xcolor}
\shorttitle{Coronal mass ejection deformation at 0.1 au}
\shortauthors{Braga et al.}
\graphicspath{{./}{figures/}}

\begin{document}

\title{Coronal mass ejection deformation at 0.1 au observed by WISPR}

\correspondingauthor{Carlos R. Braga}
\email{cbraga@gmu.edu}

\author[0000-0003-1485-9564]{Carlos R. Braga}
\affiliation{George Mason University \\
Fairfax, VA 22030 USA}

\author[0000-0002-8164-5948]{Angelos Vourlidas}
\affiliation{The Johns Hopkins University Applied Physics Laboratory \\
Laurel, MD 20723 USA}

\author[0000-0002-5068-4637]{Paulett C. Liewer}
\affiliation{Jet Propulsion Laboratory, California Institute of Technology \\
Pasadena, CA 91109 USA}

\author[0000-0003-1377-6353]{Phillip Hess}
\affiliation{Space Science Division, U.S. Naval Research Laboratory \\ 
Washington, DC 20375 USA}

\author[0000-0001-8480-947X]{Guillermo Stenborg}
\affiliation{The Johns Hopkins University Applied Physics Laboratory \\
Laurel, MD 20723 USA}

\author[0000-0002-1859-456X]{Pete Riley}
\affiliation{Predictive Science Inc. \\
San Diego, CA 92121 USA}

\begin{abstract}

Although coronal mass ejections (CMEs) resembling flux ropes generally expand self-similarly, deformations along their fronts have been reported in observations and simulations. 
We present evidence of one CME becoming deformed after a period of self-similarly expansion in the corona. The event was observed by multiple white-light imagers on January 20-22, 2021. The change in shape is evident in observations from the heliospheric imagers from the Wide-Field Imager for Solar Probe Plus (WISPR), which observe this CME for $\sim$ 44 hours.
We reconstruct the CME using forward-fitting models. In the first hours, observations are consistent with a self-similar expansion but later on the front flattens forming a dimple. Our interpretation is that the CME becomes deformed at $\sim0.1\ au$ due to differences in the background solar wind speeds. The CME expands more at higher latitudes, where the background solar wind is faster. We consider other possible causes for deformations, such as loss of coherence and slow-mode shocks. The CME deformation seems to cause a time-of-arrival error of 16 hours at $\sim 0.5\ au$. The deformation is clear only in the WISPR observations and, it thus, would have been missed by 1~AU coronagraphs.
Such deformations may help explain the time-of-arrival errors in events where only coronagraph observations are available. 
\end{abstract}

\keywords{Sun: coronal mass ejections (CMEs) --
             Sun: corona --
             Sun: solar wind }


\section{Introduction} \label{sec:intro}

Coronal mass ejections (CMEs) are giant magnetized plasma structures ejected from the Sun. Hundred of CMEs are observed every year, even in solar minimum. The CME-entrained electrons can be imaged by remote sensing instruments in white-light, such as coronagraphs and heliospheric imagers thanks to Thomson scattering \citep{Vourlidas_Howard_2006}. 
The brightness observed is primarily associated with the particle density, and the location of both the observer and the light source, but it can also give insights about their magnetic field configuration \citep{Vourlidas_2014}.

Remote sensing instruments have observed thousand of CMEs from 1 au over the last decades and they suggest that many CMEs are likely magnetic flux rope structures \citep{Vourlidas2012, Vourlidas2017}. This magnetic structure results in a circular cross section \citep{Chen2017}, that is normally associated to the dark void observed in the images. Many CMEs show a circular-like front when observed close to the Sun by remote sensing. This shape is consistent with the flux rope model, which has its apex in the central portion. 

Most CMEs expand at least approximately self-similarly in the corona \citep{Low1984, Low1987, Colaninno2006, Vourlidas2010, Subramanian2014}. A study involving 475 CMEs suggests that 65\% of them expand self-similarly at 10 solar radii \citep{Balmaceda2020}. Self-similarly is also frequently assumed in the interplanetary medium when deriving the CME Time-of-Arrival \citep{Colaninno2013, Davies2012,Davies2013, Moestl2012, Lario2020}.

Some CMEs lack a circular-like front when observed by coronagraphs or heliospheric imagers, suggesting deformation.
Many studies indicate a degree of flattening in the CME front or pancaking \citep{Kay2021, Isavnin2016,Wang2018, Riley2004, Savani2011,Odstrcil1999}. Some empirical CME propagation models consider an elliptical cross section to mimic this effect \citep{Mostl2018, Weiss2021}. 

A few CMEs have concave CME fronts, where the CME edges are further from the Sun than its central portion, in disagreement with the expectations from the flux rope model. For instance, \citet{Savani2010} observed this deformation on a CME  between 20 and 50 solar radii on November 15, 2007. They explained the deformation by the higher solar wind speed observed in the CME edges that are located at higher latitudes where the fast solar wind is predominant. \citet{Howard2012} observed 3 CME cavities transforming from concave inward (curving away from the Sun), becoming flatter, and then concave outward around $0.065\ au$ from the Sun.

Although CMEs are not commonly observed as deformed in the corona, their in situ counterparts (known as ICMEs) seem to be deformed more frequently. Multiple studies suggest that a significant fraction of the ICMEs have cross sections that are not circular but rather more complex \citep{Hu2002, Mostl2009,Chi2021,Hidalgo2016}. \citet{Nieves2018} found that 41\% of 337 ICMEs were inconsistent with the circular-cylindrical geometry expected for flux ropes \citep{Burlaga1971}. Common deformations are compression in the back side of the structure and contraction \citep{Nieves2018}. Concave and flat fronts are also observed and modelled for ICMEs. One example of deformation is the kinematically-distorted flux rope model, which assumes a flux rope with circular cross-section close to the Sun and distorted by radial propagation at the ambient solar wind. In solar-minimum-like solar wind, slow solar wind occurs at low latitude and fast wind at high latitude creating a concave ICME cross section \citep{Owens2006, Davies2021}.

When comparing studies based on ICMEs with those from CMEs, there is an apparent discrepancy in the frequency of deformations, which seems to be less common in the corona. 
We suggest that the discrepancy is because many CMEs may be deformed beyond the field-of-view of the coronagraphs.
Despite decades of CME and ICME observations, this topic was not extensively explored as the region between the corona and 1 au is not fully observed. Magnetohydrodynamic (MHD) simulations suggest that CME deformations take place. Multiple models show that CME front can become flat or concave when they are immersed into a structured solar wind with slower wind around the equatorial region \citep{Odstrcil2004, Riley2004}.

We present observational evidence of a CME deforming near the Sun observed on January 20-22, 2021, and discuss possible causes. We conclude that the structure of the background solar wind is the most likely cause of the deformation.

This paper is organized as follows. In Section \ref{sec:observations}, we describe observations available from the source region up to corona until approximately $40$ solar radii. We present evidence of CME deformation in Section \ref{sec:reconstructing}, where we reconstruct the CME by combining observations from 3 viewpoints. We also evidence differences in the CME front position at multiple latitudes in Section \ref{sec:kinematics}. Section \ref{sec:discussion} discusses the possible causes for the deformation. The implication of the deformation for the calculation of the CME ToA are explained in Section \ref{sec:implications}. Finally, Section \ref{sec:summary} has the summary and conclusions.

\section{White-light Observations}
\label{sec:observations}

\begin{figure}
    \centering
    \includegraphics[width=12cm]{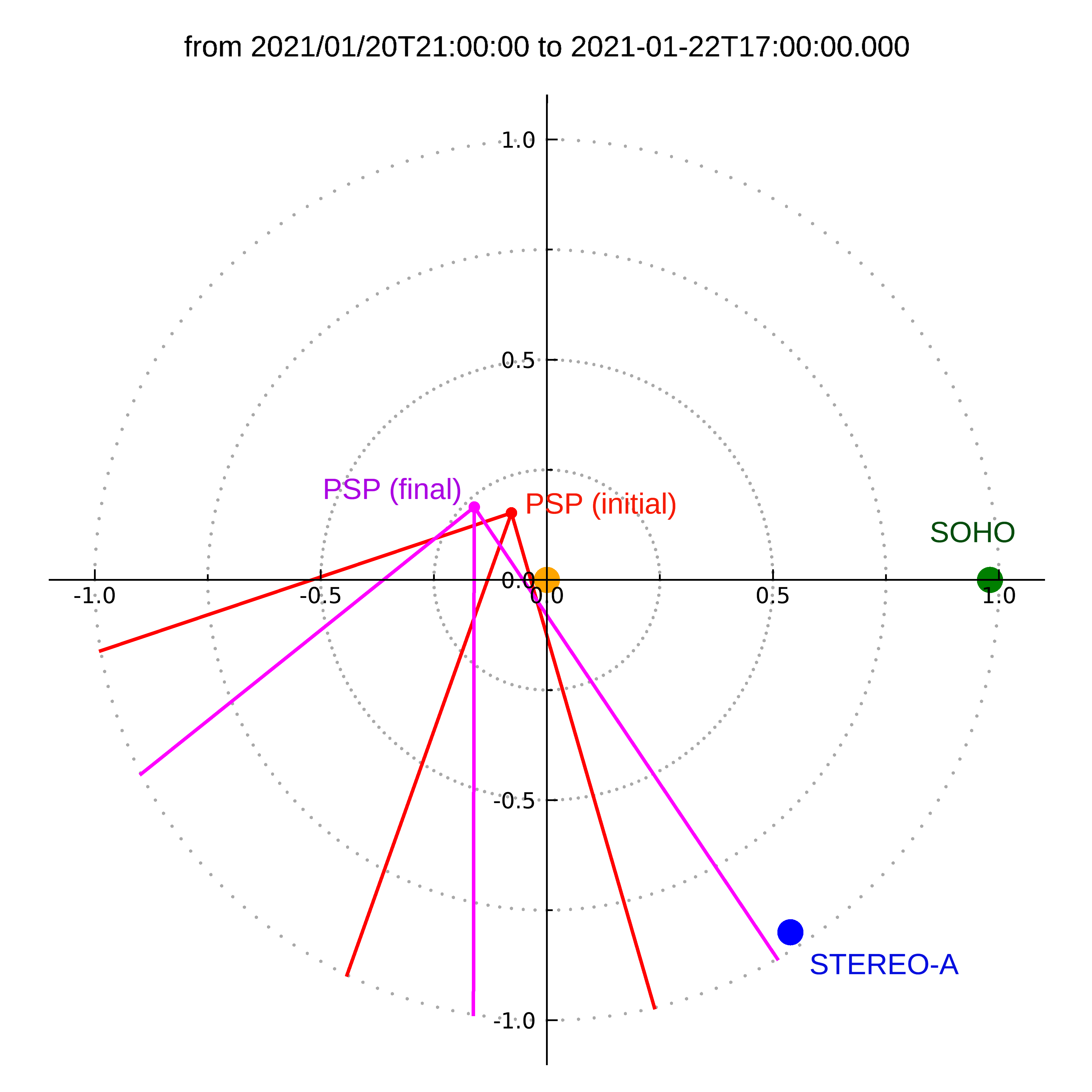}
    \caption{Position of the 3 observing spacecraft (circles). As PSP position changes significantly while WISPR observes the CME observation (from 2021/01/20 21:00 UT to 2021/01/22 17:00 UT), we indicate the spacecraft position at the first and last observations here (red and magenta circles). The lines with origins at each PSP position indicate WISPR-I and WISPR-O fields-of-view limits.}
    \label{fig:observatories}
\end{figure}

The CME under study was observed by the Wide-Field Imager for Solar Probe Plus (WISPR) \citep{Vourlidas2016}, the heliospheric image onboard Parker Solar Probe (PSP) \citep{Fox2015}. WISPR is located in the ram side of the spacecraft and has a 95$^{\circ}$ field of view (FOV) split across two telescopes: WISPR-I (inner; 13.5$^{\circ}$ to 53$^{\circ}$ elongation) and WISPR-O (outer; 50.5$^{\circ}$ to 108.5$^{\circ}$ elongation) \citep{Vourlidas2016}. The PSP locations and WISPR FOVs are indicated in Figure \ref{fig:observatories}.

WISPR observed this CME for approximately 44 hours (between 2021-01-20 $\sim$ 21:00 and 2021-01-22 $\sim$ 17:00), a few days after the 7th PSP perihelion. PSP position changed significantly during the CME observation; the solar distance increased from 0.17 au to 0.23 au while the longitudinal angle with Earth increased from 117$^{\circ}$ to 134$^{\circ}$. The PSP latitude is within $1^{\circ}$ of the ecliptic plane in the period mentioned (Figure \ref{fig:observatories}). 

White-light imaging of this CME was acquired by Sun Earth Connection Coronal and Heliospheric Investigation (SECCHI) \citep{Howard2008}, onboard Solar Terrestial Relations Observatory (STEREO) \citep{Kaiser2007} and the Large Angle Spectroscopic Coronagraph (LASCO) \citep{Brueckner1995} onboard the Solar and Heliospheric Observatory (SOHO) \citep{Domingo1995}.
SOHO's and STEREO'S locations are also shown in Figure \ref{fig:observatories}. This CME, propagating away from Earth, was not observed by SECCHI's heliospheric imagers, which are oriented to observe Earth-directed CMEs.

\begin{figure}
\begin{interactive}{animation}{cmeO7_inner_crop.mpg}
\includegraphics[width=10cm]{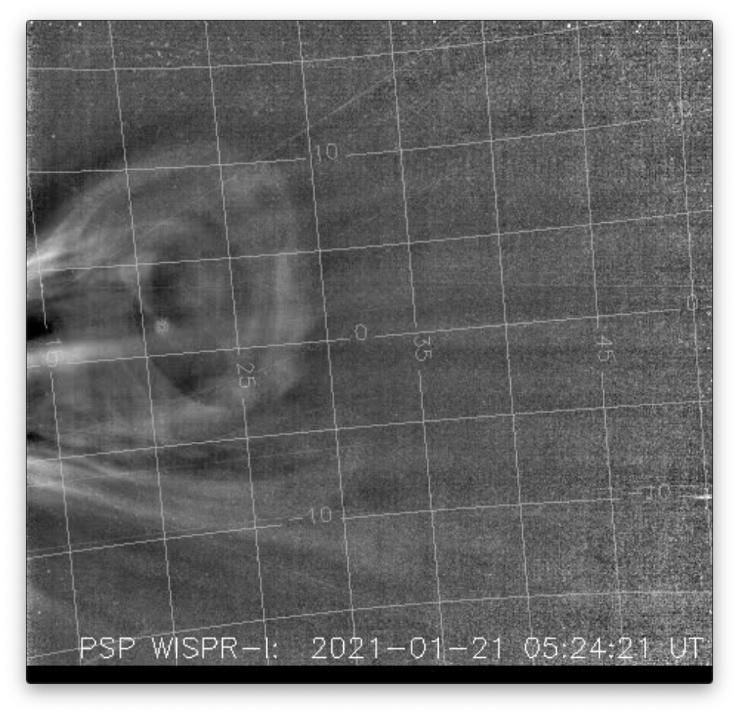}
\end{interactive}
\caption{CME observations by PSP/WISPR-I on 2021/01/22. The CME moves from left to right in the images while it advances. The Sun is located outside the field-of-view towards the left, at the coordinate system origin. The horizontal grid line at $0^{\circ}$ indicates PSP's orbit plane. The grids are given in degrees in both dimensions. An animation is available in the online journal. 
\label{fig:video_wispr_i}}
\end{figure}

\begin{figure}
\begin{interactive}{animation}{cmeO7_outer_crop.mpg}
\includegraphics[width=10cm]{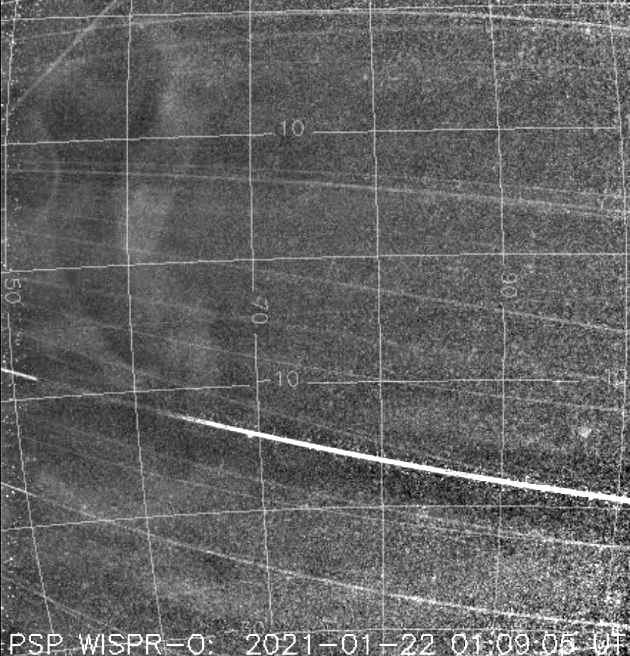}
\end{interactive}
\caption{CME observations by PSP/WISPR-O on 2021/01/22. The CME moves from left to right in the images while it advances. The Sun is located outside the field-of-view towards the left, at the coordinate system origin. The horizontal grid line at $0^{\circ}$ indicates PSP's orbit plane. The grids are given in degrees in both dimensions. An animation is available in the online journal.
\label{fig:video_wispr_o}}
\end{figure}

The CME front, as observed by WISPR, clearly changes shape with time (Figures \ref{fig:video_wispr_i} and \ref{fig:video_wispr_o}). It is approximately circular in the first hours of 2021-01-21, resembling most CMEs observed by coronagraphs and heliospheric imagers. The front flattens a few hours later, particularly after 12:00 UT. After ~18:00 UT and while in the WISPR-O FOV, the edges of the CME overtake the central portion. We call this front aspect concave. 
This shape has not been observed previously by WISPR \citep{Howard2019, Hess2020, Braga2021, Liewer2021, Wood2021}. However, it resembles the front shape observed by \citet{Savani2010} in the SECCHI heliospheric imager.

We use two independent approaches to assess the deformation: (i) reconstruct the CME using forward-modeling (Section \ref{sec:reconstructing}) and (ii) determine the kinematics of multiple CME front points (Section \ref{sec:kinematics}).

\subsection{Observations from other sources and timeline}
\label{sec:obser_other}

\begin{figure}[ht!]
    \centering
    \includegraphics[width=16cm]{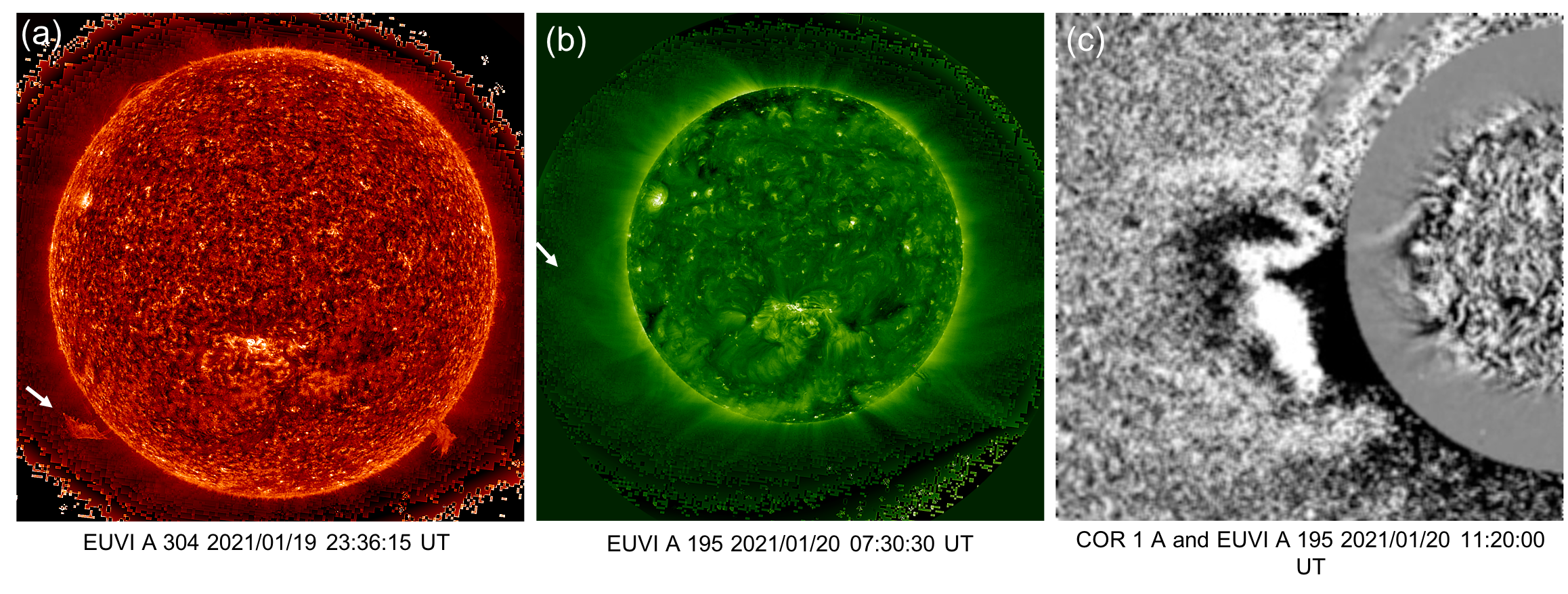}
    \caption{Multiple observations from STEREO-A indicating the solar origins of the CME: (a) EUVI 304 showing the prominence; (b) EUVI 195; and (c) EUVI and COR1 combined.
    \label{fig:euvi_cor1}}
\end{figure}

The origin of this CME was captured by the SECCHI suite on STEREO-A (STA). The eruption of the CME’s prominence is seen in the Extreme UltraViolet Imager onboard STA (EUVI-A) 304 on 2021-01-19 22 UT (Figure \ref{fig:euvi_cor1}, left). The prominence is seen well south of the solar equator and the heliospheric current sheet location  (about $10^{\circ}$ north of the solar equator at the east limb as seen from STA according to LASCO/C3 synoptic maps used in \cite{Liewer2022}). The asymmetric CME expansion towards the solar equator and the heliospheric current sheet are seen in EUVI-A 195 (Figure \ref{fig:euvi_cor1}, center) and, later, in the composite difference of COR1-A and EUVI-A 195  (Figure \ref{fig:euvi_cor1}, right). The asymmetric expansion towards the solar equator is a common characteristic of CME originating from weak magnetic field regions and it is driven by the force balance between the ambient coronal field and the CME plasma and magnetic pressures \citep[e.g.][]{Byrne2010,Liewer2015}. By about 20:00 UT, the CME has fully emerged from the corona in the shape of a classic flux rope CME. At this point, the CME appears to expand symmetrically.

As the prominence and the CME are located in the east limb as observed from STA, we do not have extreme ultraviolet observations from the Earth viewpoint.

\section{Reconstructing the CME}
\label{sec:reconstructing}

The CME 3D morphology cannot be quantified solely from WISPR observations. PSP has a highly elliptical orbit. As a result, the distance and angle between the CME and observer also change quickly. For instance, the CME size in the WISPR fov should increase when the distance to the CME decreases. A second reason is that PSP could be inside the CME so that WISPR images only a fraction of the structure (see, e.g., \citet{Liewer2019}). Another complication is that CMEs are optically thin structures. Thus, we observe brightness integrated over multiple layers along the line of sight. A given free electron in the corona produces a different brightness level due to the Thomson scattering depending on its angle with the observer. The brightness is maximum when an electron in the Sun-electron direction is perpendicular to the observer-electron line. For all these reasons, the CME reconstruction is critical for WISPR observations.

To derive the CME 3D morphology, we reconstruct the CME  using forward-modeling. We use the Graduated Cylindrical Shell model \citep{Thernisien2006a, Thernisien2011}. This is an empirical model meant to reproduce a coronal mass ejection that follows a flux rope structure without any deformation. The model is formed by two conical legs connected to a curved front reminiscent of a torus. This front has a circular cross-section that increases with height. Because of its shape, this is sometimes called the “croissant” CME model.
Dimensions, position, direction, and orientation of the structure are adjusted to reproduce the observations. 

We reconstruct the CME in order to confirm that the concave front observed is indeed a CME deformation rather than a projection effect. A second goal is understanding when and where the deformation begins.

\begin{figure}
    \centering
    \includegraphics[width=18cm]{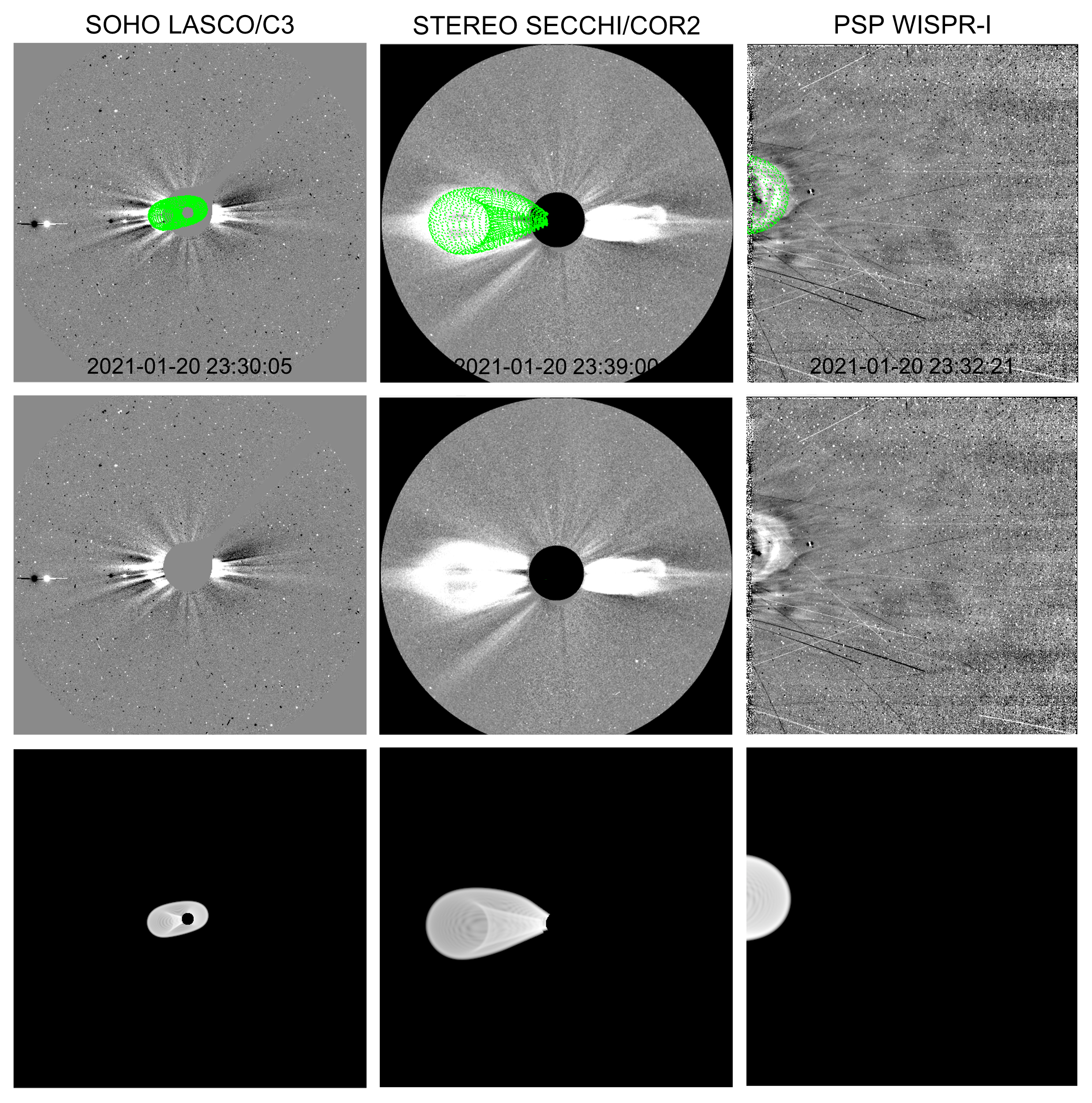}
    \caption{The CME observations on 2021/01/20 around 23:30 (central row), its GCS modelling projected over each observation (upper row), and its synthetic images (lower row). }
    \label{fig:reconstruction23}
\end{figure}

\begin{figure}
    \centering
    \includegraphics[width=18cm]{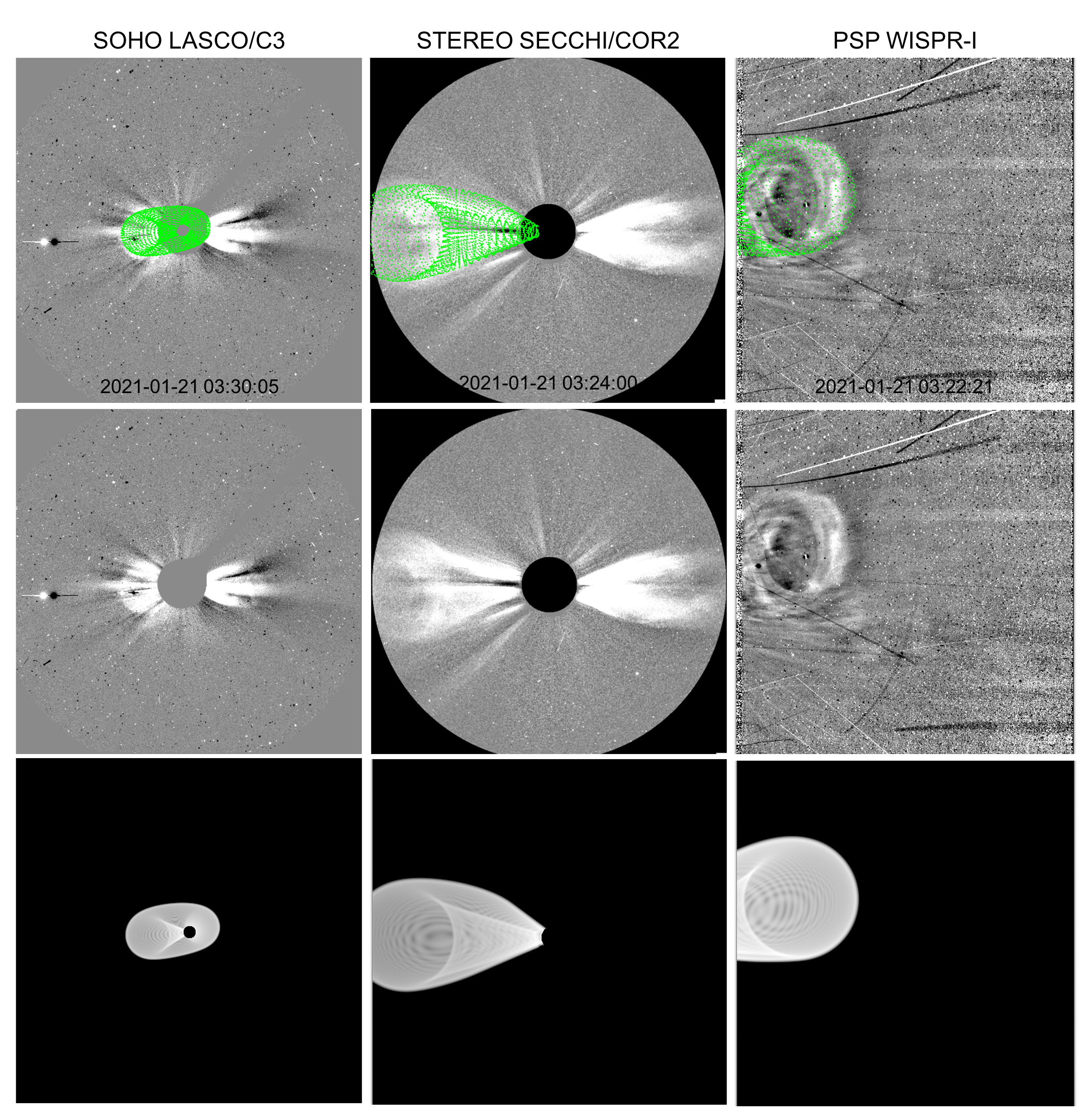}
    \caption{The CME observations on 2021/01/21 around 03:30 (central row), its GCS modelling projected over each observation (upper row), and its synthetic images (lower row). }
    \label{fig:reconstruction3}
\end{figure}

We illustrate the reconstruction in two different times:  2021-01-20 at around 23:30 in Figure \ref{fig:reconstruction23}, and 2021-01-21 at around 03:00 in Figure \ref{fig:reconstruction3}. In both cases, we have observations from WISPR-I, SECCHI, and LASCO. The green wire over each image indicates the CME reconstruction. In both times, WISPR observes a circular front that can be reconstructed well with the GCS model. 
We also derive a synthetic image with the projection of the model in each instrument. These synthetic images take into account the GCS geometry, an assumed free electron density, and the brightness that each electron produces due to the Thomson scattering. We derive the intensity of each pixel in the synthetic image by integrating the brightness along the line of sight. 
Following the GCS model, the electron population is placed in the outer shell of the structure, keeping the interior hollow. The lower rows in Figures \ref{fig:reconstruction3} and \ref{fig:reconstruction23} show the synthetic images derived for the 3 instruments.

\begin{figure}
    \centering
    \includegraphics[width=19cm]{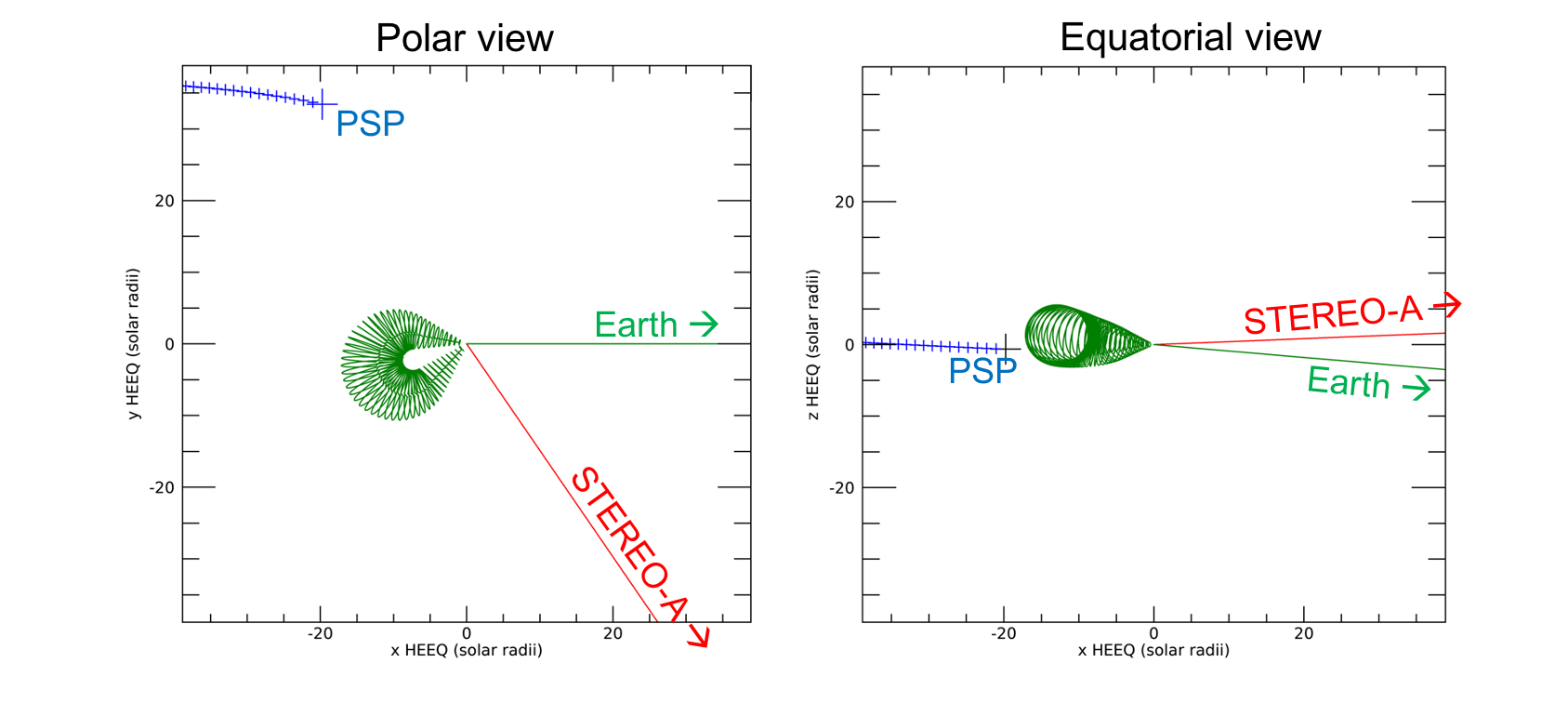}
    \caption{The CME reconstruction at 2021/01/21 03:30 UT. The left panel has the CME projection in the solar equatorial plane, resembling a polar view. The right panel shows the plane formed by the solar rotation axis and Earth's direction, corresponding to a view from the solar equator in quadrature with Earth. The larger blue '+' represents PSP position at 03:30, and the smaller ones indicate the position in the following hours. The red and green lines indicate the directions to STEREO-A and Earth.}
    \label{fig:cme_3d_view}
\end{figure}

The CME height is $11.5\ R_{\odot}$ in the first time, and $17.5\ R_{\odot}$ in the second. The remaining GCS model parameters are identical in both cases: half-angle $\alpha=31^{\circ}$, longitude $\phi=212^{\circ}$, latitude $\theta=6^{\circ}$ (both in Carrington coordinates), tilt angle $\gamma=-8^{\circ}$, and aspect ratio $\kappa=0.3$. This means that the CME extends $36^\circ$ in latitude, $62^\circ$ in longitude, and it is almost parallel to the solar equator. It propagates away from Earth ($200^\circ$ Stonyhurst longitude), and a few degrees above the solar equator ($6^\circ$ Stonyhurst latitude). We estimate a $\sim 300\ km/s$ speed in this period. The CME reconstruction at 2021-01-21 03:30 UT is shown in Figure \ref{fig:cme_3d_view}. At this time, its apex is at $17.5\ R_{\odot}$ and $30-40\ R_{\odot}$ from PSP (represented by a plus sign in the Figure \ref{fig:cme_3d_view}. The CME is not directed to any spacecraft. Both STEREO-A and SOHO are located outside the region shown, and the red and green arrows indicate their directions. 

The GCS model reproduces well the observations in both Figures \ref{fig:reconstruction23} and \ref{fig:reconstruction3}. We also reconstruct the CME at intermediate times with the same model models parameters, apart from height. In these times, we have a good agreement between observations and models. These results indicate that the CME is expanding self-similarly between 2021/01/20 23:30 and 2021/01/21 03:30 without any deformation. 

\begin{figure}
    \centering
    \includegraphics[width=18cm]{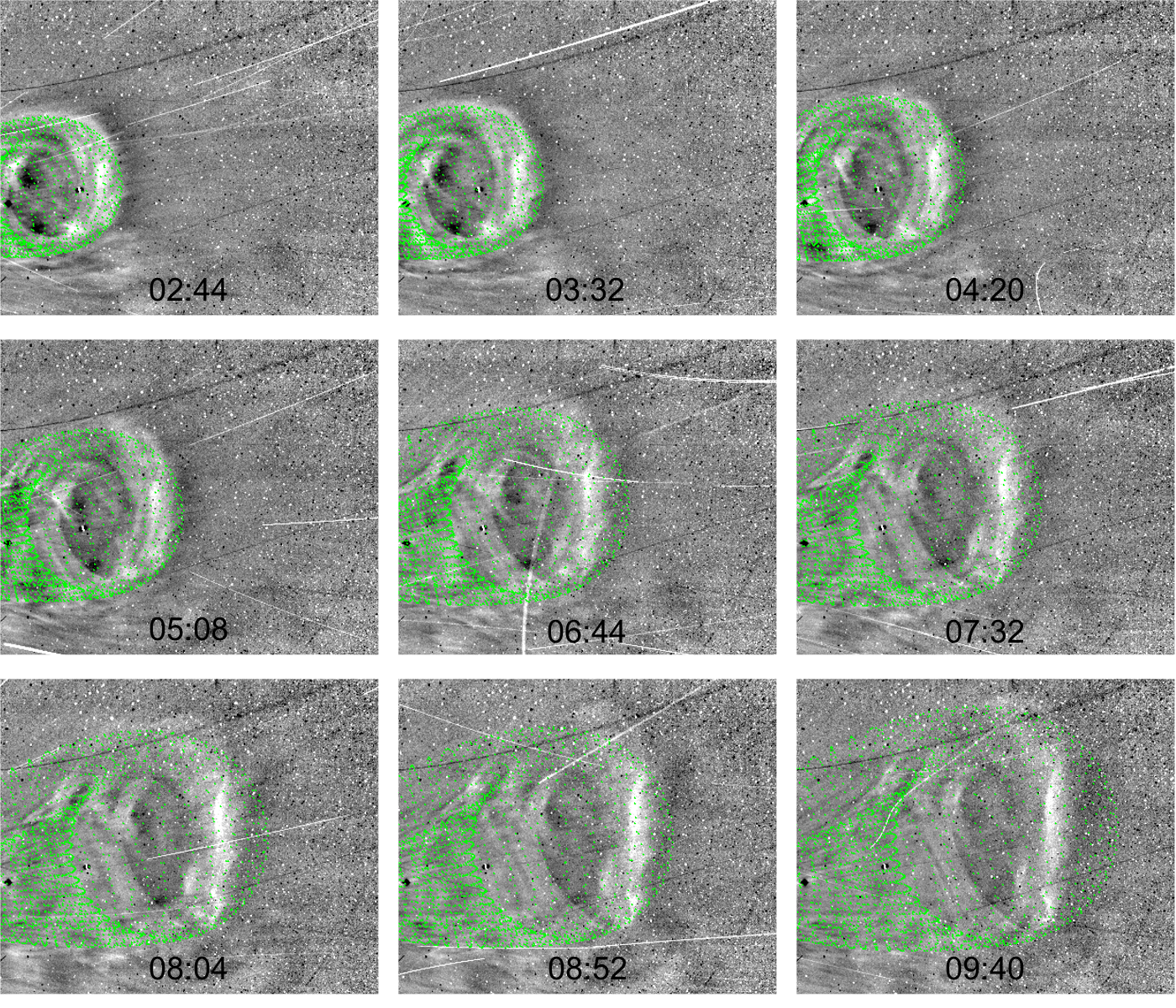}
    \caption{The CME reconstruction using the GCS model on 2021-01-21 at multiple times projected over WISPR-I images (cropped to the CME region). We keep all model parameters unchanged, except for height. Deformation becomes evident as time progresses, and the model becomes unable to reconstruct the CME observations.}
    \label{fig:cme_distorted}
\end{figure}

In the following hours, however, self-similarity cannot be assumed. If we keep all model parameters unchanged except height, the GCS model does not reproduce the WISPR-I observations after 2021-01-21 06:00 UT (Figure \ref{fig:cme_distorted}). 
To our assessment, the model does not reproduce these observations even if we relax self-similarity. Although we experimented with an extensive list of parameters (angular width, longitude, rotation, latitude, etc), we were unable to reproduce the observation with the model. Conversely, the GCS model seems to agree with the LASCO/C3 observations from the same period, which have a circular-shaped front. We do not have COR2 observations for the entire period shown in Figure \ref{fig:cme_distorted}, as the CME front exits the COR2 FOV. The heliospheric imagers from STEREO-A/SECCHI do not observe this CME as it is directed away from Earth. Thus, only LASCO/C3 and WISPR-I observations are available in most times shown in Figure \ref{fig:cme_distorted}.   
The CME exits the WISPR-I FOV on January 21 at around 19:00 UT and continues into WISPR-O. Again, the GCS model cannot account for the observed CME front shape, regardless of the parameters we use.

In summary, we conclude that this CME first expands self-similarly, but then begins to deform. This process becomes evident between $17\ R_{\odot}$ ($0.08\ au$) and $32\ R_{\odot}$ ($0.12\ au$) in the WISPR FOV. It seems to be gradual and continuous from this point on, leading to evident changes in the CME front shape from convex to concave. 

\subsection{Another CME}

Here we briefly assess another CME observed during the same time. Both LASCO C2/C3 and COR2 observe this event in the west limb (upper central and left panels in Figures \ref{fig:reconstruction23} and \ref{fig:reconstruction3}). After leaving the COR2 FOV, the CME propagates through the HI-1 FOV, which covers the Sun-Earth line between $4^{\circ}$ and $24^{\circ}$ of elongation. WISPR does not observed this event as it is outside its FOV. 

These observations indicate that this second event propagates approximately in the opposite direction than the CME reconstructed. Thus, we do not expect interaction between these CMEs, and deformation is unlikely to be explained by interaction.    

\section{The CME kinematics}
\label{sec:kinematics}

Due to the front deformation, we avoid relying on the graduated cylindrical shell model to derive the CME position. Instead, we use an alternative method to extract the CME kinematics that does not require a geometric model for the CME. 

As the PSP-CME distance varies significantly while WISPR observes the CME, we need to take into account the effect of spacecraft motion. This effect is negligible for 1 AU STEREO observations but not for PSP, where the longitude and solar distance change significantly in less than a day. Thus, we cannot use most methods adopted to STEREO heliospheric imagers here, such as fixed-$\phi$ or harmonic mean (see discussion in \cite{Liewer2019} for more details).

We track a point at the CME front. The kinematics of the point are derived by calculating the 4 free parameters of the model: the point's spherical coordinates in the first image, and its initial velocity. We consider that the point propagates radially away from the Sun with constant speed to limit the number of free parameters (Figure \ref{fig:CMEpointpos}). This method is described and applied to CME kinematics in \cite{Braga2021}. It is a variation of the original method from \cite{Liewer2019} and adds the option to include acceleration and changes in longitude as free parameters.

We select a pixel with the feature of interest in each WISPR image, and extract its angular coordinates in the image plane ($\gamma$, $\beta$). $\gamma$ is that the elongation angle on the orbit plane. $\beta$ is the angle out of the orbit plane. Both angles have vertex on the spacecraft. 

As the spacecraft position changes quickly over time, the transformation between pixels and angular coordinates is time-dependent. 
The determination of $\gamma$ and $\beta$ requires the spacecraft location and attitude, as well as the camera projection and distortion effects. Following \cite{Braga2021} and \cite{Liewer2019}, we compute this angles using the routine \texttt{wispr\textunderscore camera\textunderscore coords.pro} \normalfont that is part of the WISPR library on SolarSoft \citep{Freeland1998}.

Other way to derive $\beta$ and $\gamma$ is using the geometric relation between the spacecraft and CME positions shown in Figure \ref{fig:CMEpointpos}. Following \cite{Liewer2019}, we have the following expressions: 

\begin{figure}
    \centering
    \includegraphics{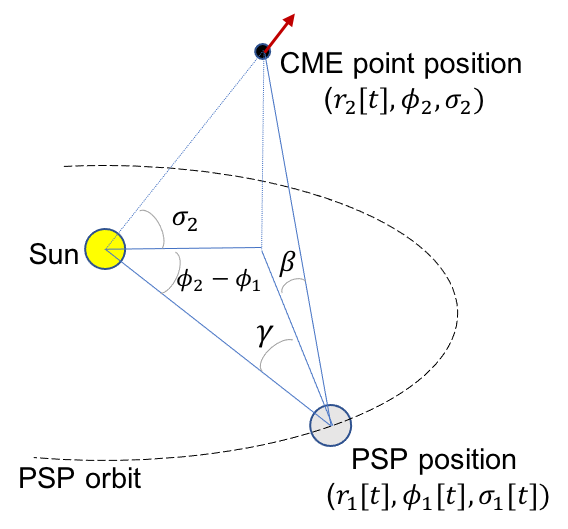}
    \caption{Determination of a CME point position ($r_{2}[t],\phi_{2},\sigma_{2}$) using the spacecraft coordinates ($r_{1}[t],\phi_{1}[t],\sigma_{1}[t]$) and extracting the angles $\beta$ and $\gamma$ from the images.}
    \label{fig:CMEpointpos}
\end{figure}

\begin{figure}
    \centering
    \includegraphics[width=10cm]{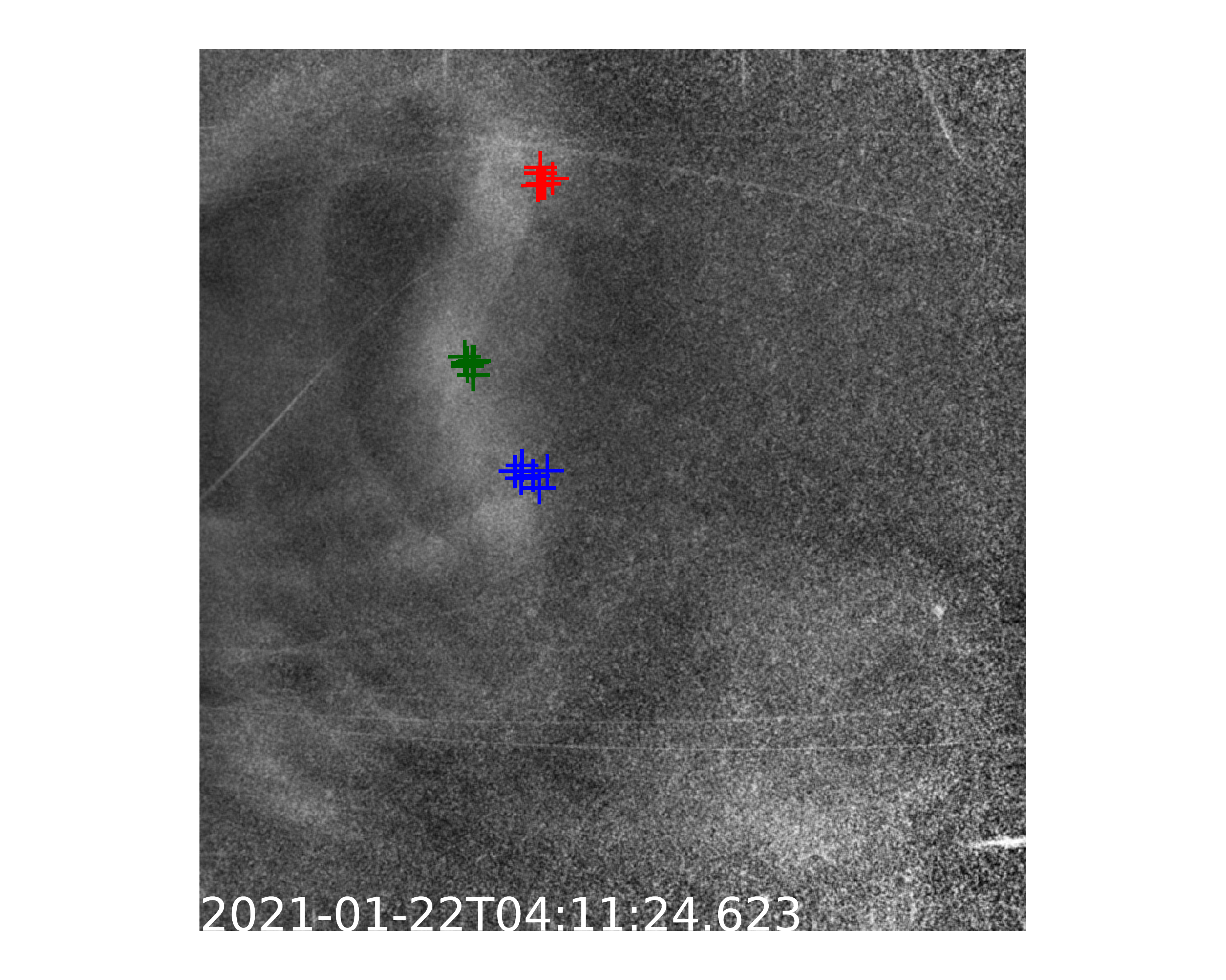}
    \caption{We calculate the position and speed of 3 different CME front portions on WISPR-O observations: the furthest points in the northern and southern parts of the CME front (red and blue crosses), and the latitude where the front looks closer to the Sun (green crosses). 
    As we repeat the front identification by eye 6 times for each portion to estimate errors in the kinematics, we have 6 crosses for each color.}
    \label{fig:wispr_o_3_points_location}
\end{figure}

\begin{equation}
tan\,(\beta[t]) = \frac{tan\,(\sigma_{2})\ sin\,(\gamma[t])}{sin\,(\phi_2-\phi_1[t])}
\label{eq:tanbeta}
\end{equation}

\begin{equation}
cotan\,(\gamma[t]) = \frac{r_{1}[t]-r_{2}[t]\ cos\,(\sigma_{2})\ cos\,(\phi_{2}-\phi_{1}[t])}{r_{2}[t]\ cos\,(\sigma_{2})\ sin\,(\phi_{2}-\phi_{1}[t])}
\label{eq:cotangamma}
\end{equation}

where the subscript $2$ indicates the CME-point spherical coordinates: radial distance $r_{2}$ as a function of time $t$, longitude $\phi_{2}$, and latitude $\sigma_{2}$. Both $\phi_{2}$ and $\sigma_{2}$ are given in spacecraft orbit coordinates. The subscript $1$ indicates the spacecraft coordinates, which are known. 

We determine the model parameters by selecting the one that results in minimum residual between $\beta[t]$, $\gamma[t]$ extracted from the images and $\beta[t]$, $\gamma[t]$ calculated using Equations \ref{eq:tanbeta} and \ref{eq:cotangamma}. 

We applied this method to 3 points in the CME front shown in Figure \ref{fig:wispr_o_3_points_location}: the lowest point in the central portion, and the furthest points in the southern and northern parts of the CME. We select these points because we believe they characterize the deformation. Here we use only WISPR-O observations because the deformation in WISPR-I is not clear enough in sufficient number of frames. 

We repeat our measurements 6 times to estimate the error. In each repetition, we identify the feature of interest by eye in the WISPR image sequence, and use this series of points to find the best parameter set. We use the 6 $r_{2}[t]$ profiles derived for each point to calculate its median and standard derivation. We understand that this error estimate is associated both to the feature extraction and to the fit method. We use the median as the result, and the standard deviation as error bars. 

\begin{figure}
    \centering
    \includegraphics[width=18cm]{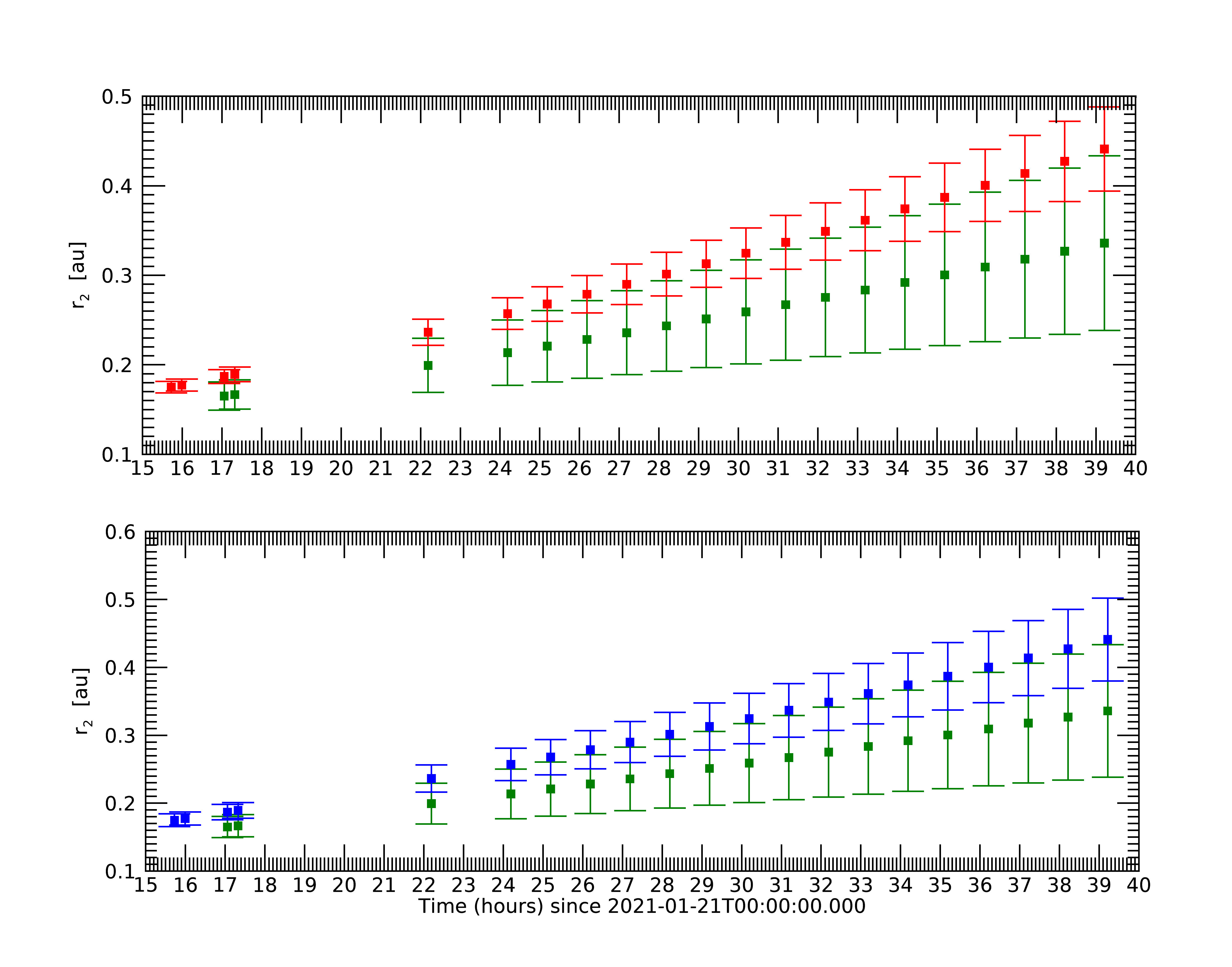}
    \caption{Position of the CME front at 3 different point in its front: northern and southern leading edges (red and blue, respectively) and the central valley (green).}
    \label{fig:position}
\end{figure}

The positions for the three points as a function of time are shown in Figure \ref{fig:position}. The southern and northern points have similar solar distances, and both are ahead of the central point. As the error bars exceed these differences, this result does not irrefutably confirm that the CME central point is closer to the Sun. 
We realize that the error bars we get for $r_{2}$ are considerably higher to those in \citet{Braga2021}. One possible explanation for the higher error is the instrument. Here we use only WISPR-O observations, and $50^{\circ}< \gamma[t]< 80^{\circ}$. \citet{Braga2021} used only WISPR-I resulting in $\gamma< 20^{\circ}$. A second possible explanation can be associated with $\phi_2[t]-\phi_1[t]$, which is higher than in \citet{Braga2021}. Our assessment suggests that $r_2$ errors are proportional to $\phi_2[t]-\phi_1[t]$. 

We summarize the  kinematic parameters of the 3 CME points in Table \ref{tab:points_kinematics}. Based on our reconstruction, the latitudes and longitudes correspond to the CME portion closer to PSP.
We believe that this particular CME portion is reflected in the kinematics because we tracked the leading edge observed by WISPR. From Figure \ref{fig:cme_3d_view}, this CME portion is more likely to correspond to the leading edge. 
In addition, this portion is closer to the Thomson sphere, and it produces the highest brightness in the images when compared to the remaining CME portions.  

\begin{table}[]
    \centering
    \begin{tabular}{c|c|c|c}
         Point &  Northern & Central & Southern \\
         Carrington latitude ($^\circ$)  & $13\pm1$ & $3\pm1$ & $-5\pm0$\\
         Carrington longitude ($^\circ$) & $198\pm5$ & $194\pm11$  & $189\pm5$\\
         Speed  ($km\ s^{-1}$)          &   $365\pm 50$ & $265\pm107$ & $365\pm63$\\
         Position at 2021-01-21T17:03:24 ($R_{\odot}$) & $40\pm2$ & $36\pm3$ & $40\pm2$\\
         Position at 2021-01-22T15:12:52 ($R_{\odot}$) & $95\pm21$ & $72\pm21$ & $95\pm13$\\
    \end{tabular}
    \caption{Comparison of the 3  points of interest in the CME front: the central dimple, and the northern and southern leading edges.}
    \label{tab:points_kinematics}
\end{table}

\subsection{Comparing of Height-time measurements with reconstruction}

The overlapping of the error bars in Figure \ref{fig:position} provides only a marginal verification for deformation. We can address the deformation in another way. We ignore the CME dimple and reconstruct the CME using the undisturbed boundaries (along the northern and southern flanks) as guides. We then compare the location of the reconstructed front to the measured height to assess the degree of deformations.   

We perform the reconstruction with WISPR-O observations using the same parameters derived from the reconstruction  at 2021/01/21 03:30 (Figure \ref{fig:reconstruction3}) except for the height, which we increase to fit the leading edges of the CME at the flanks. By doing so, we are assuming that the CME expands self-similarly. We do not make any assumption about constant speed, but our reconstruction suggests that this is the case when we take the height as a function of time. By doing so we get $\sim340\ km/s$.

\begin{figure}
    \centering
    \includegraphics[width=18cm]{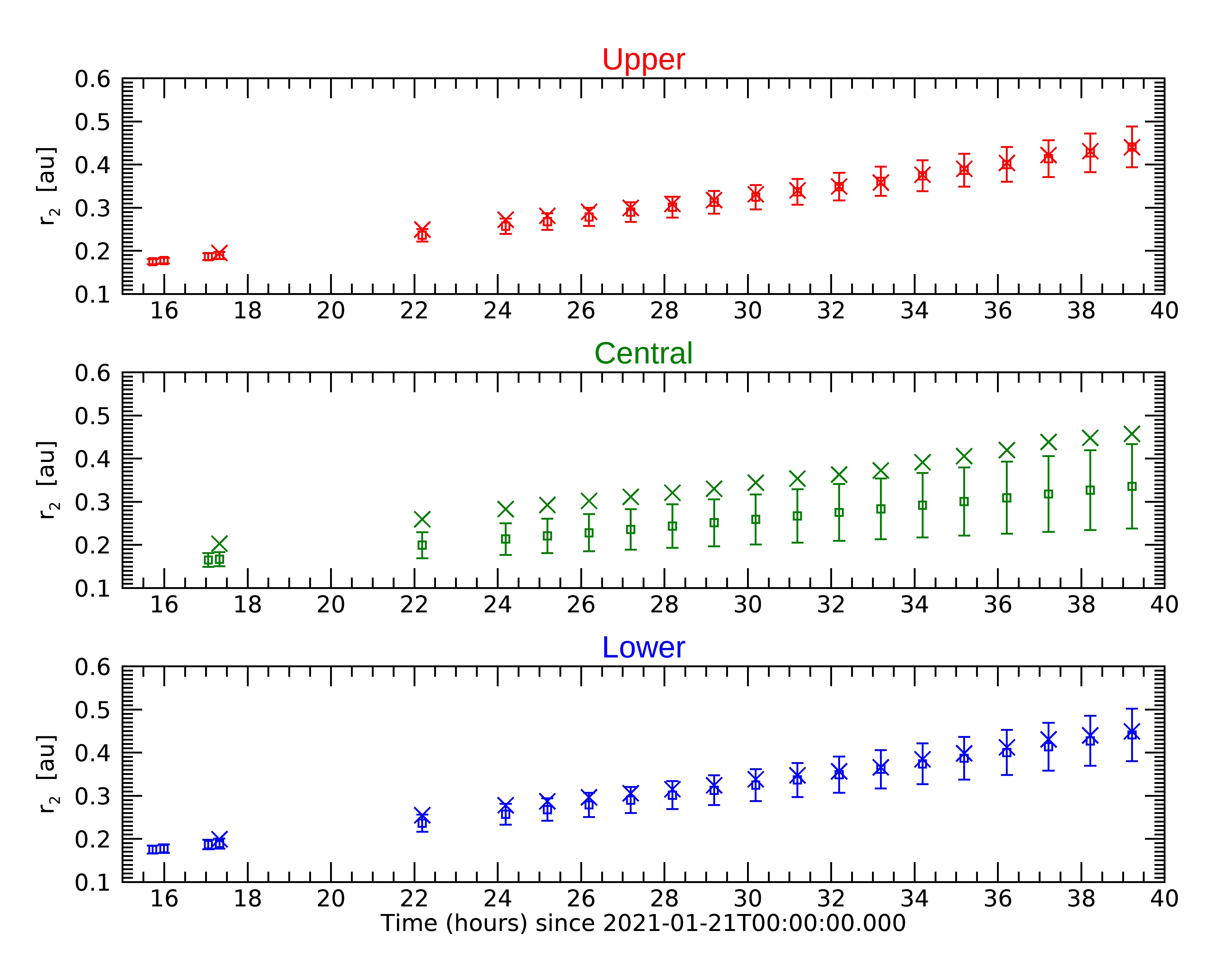}
    \caption{Comparison CME reconstruction (crosses) with the point-feature kinematics (square and the error bars) on WISPR-O fov. The model used for reconstruction is the "croissant" CME model that does not assume any deformation. As the CME deformation is obvious, we fit the model only the northern and southern leading edges, ignoring the central portion. As the reconstruction agrees well for the upper and southern leading edges, the difference between the reconstruction and the point-feature kinematics in the central portion indicates the deformation of the CME in the central latitudinal range.}
    \label{fig:position_reconst_kinematics}
\end{figure}

The comparison between the 3 height-time points and the reconstruction is shown in Figure \ref{fig:position_reconst_kinematics}. We find a very good agreement between the 3D reconstruction and the heights of the two flank points. The agreement suggests that the assumptions used to derive the height-time locations (constant speed and direction of propagation) are realistic for the period shown in Figure \ref{fig:position_reconst_kinematics}. 

In contrast to the flanks, there is a significant discrepancy between the reconstruction and height-time measurements for the CME dimple (Figure \ref{fig:position_reconst_kinematics}, middle panel). The reconstruction suggests that the CME front is deformed by $17\pm8\%$ (see details in Section \ref{sec:implications}).  

The comparison between the two techniques in Figure \ref{fig:position_reconst_kinematics} provides strong support for CME deformation. As we kept all other reconstruction parameters, except height, unchanged, we conclude that the CME expands self-similarly overall, except around the dimpled region. 

\section{Origins of the CME deformation}
\label{sec:discussion}

Here we discuss several possible origins for the CME deformation. In Section \ref{sec:background_solar_wind_models}, we investigate if the differences in the solar wind are able to explain the CME deformation. To do so, we use MHD models. 
In the remaining sections, we present other, more speculative drivers for the CME deformation: loss of coherence (Section \ref{sec:loss_coherence}) and slow shocks (Section \ref{sec:slow shocks}).

\subsection{Background solar wind speed profile}
\label{sec:background_solar_wind_models}

As introduced in Section \ref{sec:intro}, deformations in the CME front are sometimes explained by the differences in the background solar wind speed. Since the CME dimple is located close to the solar equator, where  solar wind speeds tend to be lower, we investigate here the CME background solar wind speed. 
We do not have in situ measurement from from the solar wind speed in the region around the CME as PSP longitude is far from the CME (Figure \ref{fig:observatories}). Therefore, we rely on MHD models to assess the ambient solar wind configuration. 

We use Predictive Science Inc.'s CORona-HELiosphere (CORHEL) model framework, which includes various versions of coronal and heliospheric codes called MAS (Magnetohydrodynamic Algorithm outside a Sphere). All of them solve the usual set of resistive MHD equations on a nonuniform mesh in spherical coordinates. The modeling region is separated into coronal and heliospheric domains, with the boundary typically lying at 30 Rs. We consider both MAS polytropic (MASP) solutions, where the energy equation is approximated by a simple equation of state and MAS thermodynamic (MAST) solutions, where energy transport processes are considered, albeit in a semi-empirical way. While these models provide excellent solutions for the structure of the coronal magnetic field, to varying degrees, and for different reasons, they do not reproduce accurate speed and density variations. 

MASP solves the usual set of MHD equations in spherical coordinates with the energy equation being approximated by a simple adiabatic approximation. The model has all energy source terms set to zero and this requires a polytropic index of 1.05 to reflect the near-isothermal nature of the corona, and 1.5 in the solar wind. The code reproduces the structure of the magnetic field reasonably well, but fails to generate solutions with sufficient variation in solar wind speed or densities.  This limitation is overcome using an empirically-based approach, the ``Distance from the Coronal Hole Boundary'' (DCHB), to specify the solar wind speed at the inner boundary of the heliospheric code \citep{riley01a}. Although semi-empirical, previous studies \citep{Riley2015} suggest that MASP produces results that match 1 au observations as good as those using Wang-Sheely-Arge (WSA) methodology for describing the solar wind speed at the inner boundary of the heliospheric model \citep{Arge2003}.

MAST replaces the polytropic assumption with an empirically-based treatment of energy transport processes (radiation losses, heat flux, and coronal heating) in the corona \citep{Lionello2001, Lionello2009}, and the polytropic index is 5/3. This model focused on improving the density and temperature structure in the solar corona through comparisons with EUV and X-RAY images from a variety of measurements. Thus, MAST also implements the DCHB approximation to derive the heliospheric boundary conditions from the thermodynamic solution. For this reason, the model is not strictly thermodynamic but rather ``semi-empirical thermodynamic''.

In this study, we analyze speeds and densities at various distances from the Sun, and from different model solutions. These are used to provide at least a qualitative measure of the uncertainty in the model results. Thus, results at 20 Rs are from slices taken directly from the coronal thermodynamic solution (MAST). Results at 30 Rs are effectively the innermost radial boundary conditions from the heliospheric solution. The MASP and MAST labels refer to whether the polytropic or thermodynamic solutions were used to derive the boundary conditions using the DCHB technique. In general, given the better match of the solutions using the DCHB approach with observations \citep{Riley2015}, these are expected to be more accurate. However, since they are produced farther out in the corona than the observations we are attempting to interpret, they are not ideal. For this reason, the coronal solutions at 20 Rs may be equally valid for interpreting the observations. More details about the differences between MAST and MASP models are explained in \citet{riley21e} and references therein. Finally, we also include model results that rely on two different heating profiles (101 and 201). Although we generally believe that the latter is more accurate, we include both to provide yet another marker for the model uncertainty. 

\begin{figure}
    \centering
    \includegraphics[width=18cm]{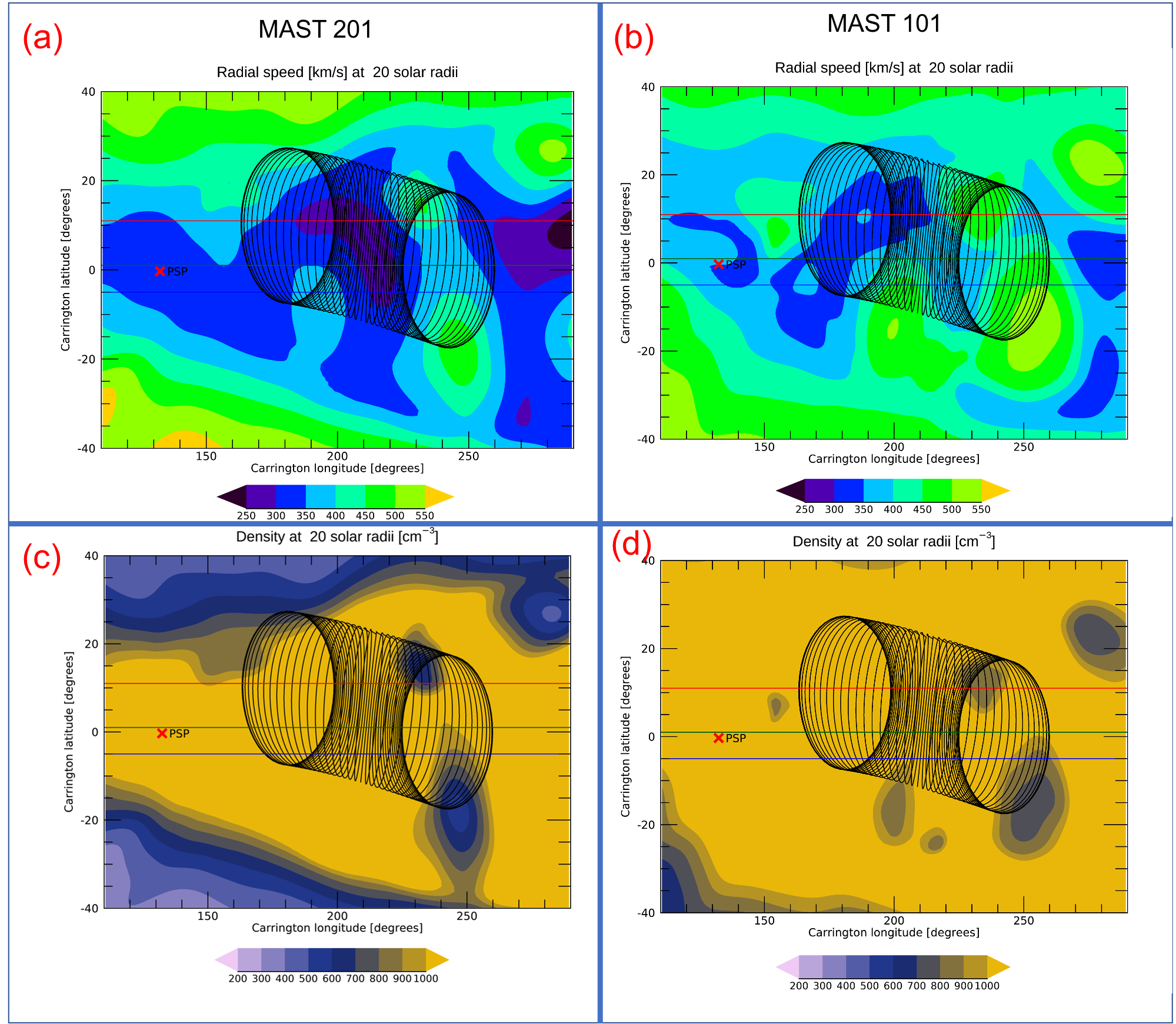}
    \caption{Radial speed and density map at 20 solar radii calculated using the CORHEL MAS MAST model. The red, green and blue horizontal lines indicate the latitude of the 3 points in the CME front of which we calculated the speed. We projected in black the CME reconstruction before the deformation.}
    \label{fig:speed_map}
\end{figure}

As the deformation begins around 20 solar radii, the radial speed derived by the CORHEL MAS MAST 101 and 201 at this spherical surface (Figure \ref{fig:speed_map}, upper row). Latitude and longitudes are both in Carrington coordinates, and these models were derived for Carrington rotation 2239. The position of the CME derived from the reconstruction (Section \ref{sec:reconstructing}) is indicated by the black lines.

We compare the solar wind radial speed from the model with the 3 points in the CME front discussed in the previous section in Section \ref{sec:kinematics}: the CME northern and southern leading edges (red and blue points) and in the central valley (green point). These latitudes are indicated in Figure \ref{fig:speed_map} by horizontal lines with the corresponding colors. The solar wind radial speed changes significantly in the CME region, ranging from 250 km/s to at least 450 km/s. Overall, the northern and southern leading edges latitudes (red and blue lines) are associated to higher solar wind speeds at least in some longitudes (particularly close to $180^{\circ}$ and $240^{\circ}$), but not in all. This suggests that CME portions close to the northern and southern leading edges are immersed in higher solar wind speed than the region close to the central dimple. This is also in agreement with the expectation that the solar wind speed is higher at higher latitudes and slower close to the equatorial region. 

We also included in Figure \ref{fig:speed_map} the density derived by the same model. Overall, the density seems to be anti-correlated with the speed as regions with low density have higher solar wind speed. This supports our explanation that the CME deformation is caused by structured solar wind. The northern and southernmost portions of the CME are associated to higher solar wind speeds when compared to the central portion. We find similar pattern in any radial distance from 20 up to the outer limit of the coronal models, at 30 solar radii. 

\begin{figure}
    \centering
    \includegraphics[width=18cm]{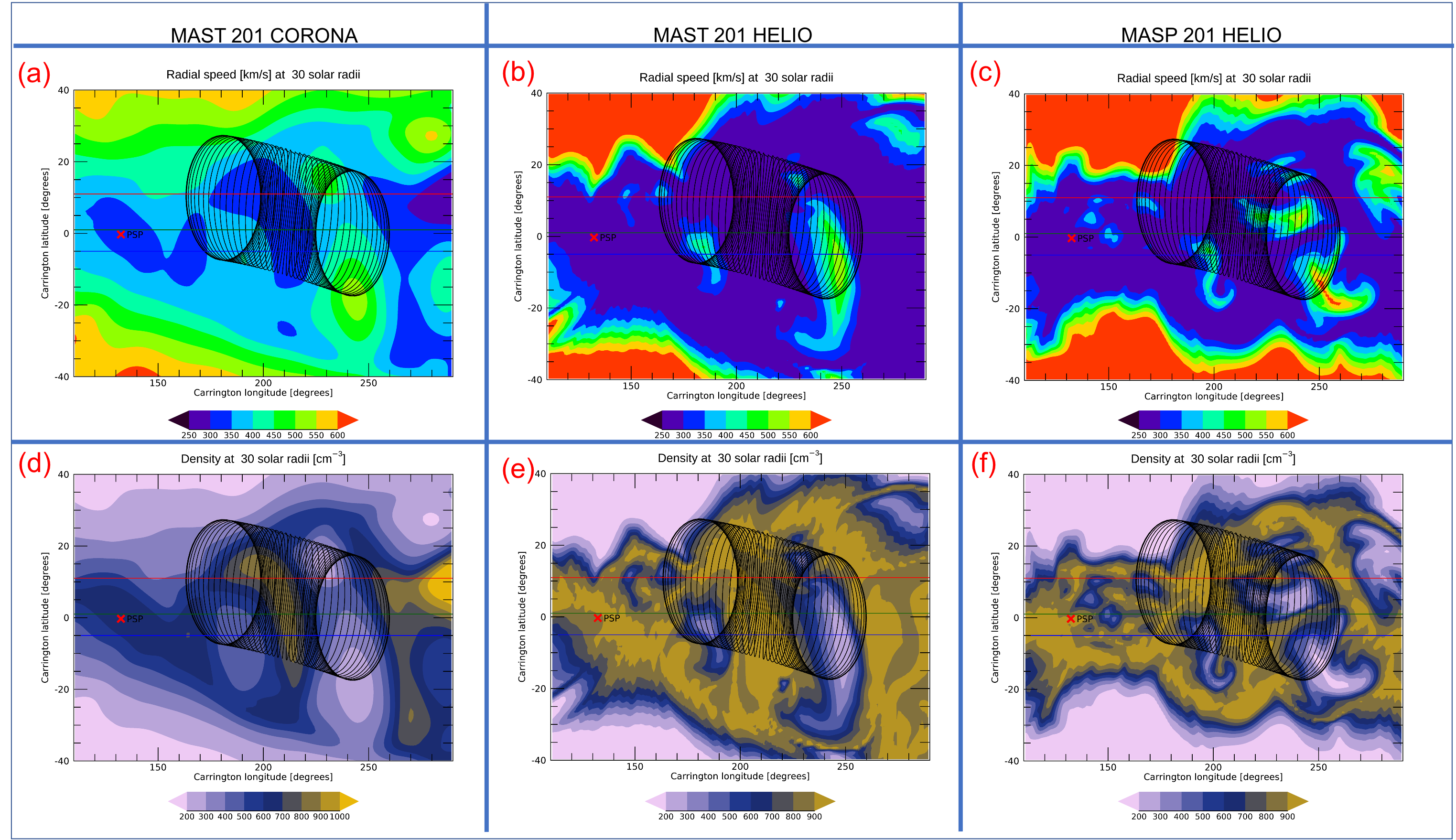}
    \caption{Radial speed and density map at 30 solar radii calculated using the CORHEL MAS MAST model. The red, green and blue horizontal lines indicate the latitude of the 3 points in the CME front of which we calculated the speed. We projected in black the CME reconstruction before the deformation.}
    \label{fig:speed_map30}
\end{figure}

We display the solar wind speed and density at 30 solar radii in Figure \ref{fig:speed_map30}. Here we include one coronal model (CORHEL MAS MAST 201) and two heliospheric (CORHEL MAST MAST 201 and MASP 201). All these 3 models have similar trends than those found at 20 solar radii.  

\subsection{CME Coherence}
\label{sec:loss_coherence}

Generally, CMEs are assumed to remain coherent as they propagate into the heliosphere. This is a basic assumption for reconstructions of CME magnetic field morphology based on in-situ measurements, for example.  
Nevertheless, \citet{Owens2017} suggested that most CMEs should loose cohesion once the separation of the CME plasma parcels exceed the local Alfvén wave speed. Since information cannot propagate within a CME faster than this speed, coherence should be lost, as different parts of the CME are subject to varying interactions with ambient structure. \citet{Owens2017} estimate that most CME lose coherence before $\sim0.3\ au$. They compared the relative speed between two CME plasma portions $v_g$ and the Alfvén speed $v_a$ within the CME. When $v_g> v_a$, they considered that coherence is lost.

Since we observe a deformation in our event, it makes senses to examine whether the \citet{Owens2017} assertion applies here. 
We estimate $v_g$ using the linear expansion rate of the CME between two times when the  deformation is starting (2021/01/21 at 03:30 and 09:30, respectively). We take the CME heights (17 and 27 solar radii) and width ($36^\circ$) from the GCS reconstruction. We find $v_g=\sim200\ km\ s^{-1}$ when comparing the separation between two points in the southern and northern flanks of the CME front. We use these points as the deformation seems to take place along the north-south direction.

To estimate the Alfvén speed, we need the number density and magnetic field along the CME front. However, we do not have in situ measurements for this event.
As a first-order approximation, we consider the Alfvén speed estimates from \citet{Owens2017} calculated for a CME with an average magnetic field and density at 1 au ($15\ nT$, $7\ cm^{-3}$). These parameters are propagated back to the CME location to estimate the Alfvén speed at the time of deformation. The assumptions are (1) constant magnetic flux within the CME, and (2) magnetic field inversely proportional to the cross-section. The density was considered proportional to the CME volume. This simple calculation results in $v_a>300\ km\ s^{-1}$ for the CME around $~0.1\ au$. 
Since $v_a> v_g$, the CME front does not satisfy the condition for loss of coherence proposed by \citet{Owens2017}. Therefore, the observed deformation is unlikely to be the result of coherence loss. We are aware of the  simplifications used to estimate $v_a$. A precise calculation of the Alfvén speed is outside the scope of this article as we are only interested in a first-order estimate. 

Although the criteria for deformation is unlikely to be satisfied for this CME, it is more feasible for faster events.
The CME studied is relatively slow ($\sim300\ km\ s^{-1}$), and some CMEs are two or three times faster. As $v_g$ is proportional to the CME speed, fast CMEs may have $v_g > v_a$. Thus, loss of coherence suggested by \citet{Owens2017} may be satisfied for fast CMEs in the corona at around $0.1\ au$.

\subsection{Is the concave front a slow-mode shock? }
\label{sec:slow shocks}

Motivated by the observations of CMEs with deformed fronts in the 1970s and 1980s by the Skylab and Solar Maximum Mission, \cite{Hundhausen1987} and references there in proposed that the concave CME fronts would be associated to slow-mode shocks while the most typical convex fronts would be associated with fast-mode shocks. 

A slow-mode shock can form when the speed in the shock frame of reference and normal to the CME front lies between the upstream sound and Alfvén speeds. Although the vast majority of measured shocks are fast-mode shocks, a few slow-mode shocks have been reported. Solar wind measurements from Helios-1 at $0.31\ au$ contain clear examples of slow shocks \citep{Richter1985, Richter1987}, as well as from other observatories \citep{Burlaga1971,Chao1970}. More recently, PSP observed slow-mode shocks at $0.31\ au$ \citep{Zhou2021}. In situ signatures of these shocks are sharp decreases in the magnetic field accompanied by an increase in density, speed and temperature when moving from upstream to downstream. Slow shocks are expected mainly in low beta plasmas in the corona, particularly if the temperature of ions is lower than the electron \citep{Hada1985}. These shocks decay faster that fast-mode shocks, and thus are unlikely to be found beyond $0.4\ au$ \citep{Grib1996}. At 1 au, a few slow-mode shocks were observed among hundreds of fast-mode (see, for example, the Harvard Smithsonian Center for Astrophysics' shock database, \url{https://www.cfa.harvard.edu/shocks/}). 

The explanation suggested by \citet{Hundhausen1987} for the concave shock is related to the angle between the shock normal and the magnetic field: in the post-shock region, this angle deflects toward (away) from the direction normal to the shock surface for slow (fast)-mode shocks. This is expected from the Rankine-Hugoniot relations.
If we consider a CME as an object where the magnetic field from the solar wind is impenetrable and we have magnetic field lines approximately radial from the Sun around the CME, the magnetic field lines can only be deflected towards the shock normal if the front is concave (see Figure 2 in \citet{Hundhausen1987}).

We examine whether slow shocks are likely between $10\ R_{\odot}$ and $20\ R_{\odot}$ by estimating the Alfvén and sound speeds. We use the magnetic field, solar wind speed, density, and temperature from the CORHEL MAS model. We estimate sound speed between $100\ km/s$ and $200\ km/s$ and the Alfvén speeds between $200\ km/s$ and $600\  km/s$. As the solar wind speed ranges from $100\ km/s$ to $200\ km/s$, it is reasonable to expect some CME with speeds between $300-400\ km/s$ to be supersonic but sub-Alfvénic. Thus, a slow-mode shock seems to be possible.

Unfortunately PSP is far from the CME in the current event, having distances of at least $20\ R_{\odot}$ (see Figure \ref{fig:cme_3d_view}), and we do not have any in situ measurement in the CME front to determine if a shock forms nor its type. 

Despite slow-mode shock observations in the references mentioned above, none of the drivers is a CME. This lack of events suggests that this type of shock is not common. Thus, associating the CME deformation with the structured background solar wind seems to be more likely than associating it with a slow-mode shock at least on a statistical basis.

\section{What are the implications of CME deformations?}
\label{sec:implications}

The CME Time-of-arrival (ToA) is a key element for Space Weather forecasting. CMEs can produce intense geomagnetic storms \citep{Gosling1993}, and they are frequently associated with acceleration of the energetic particles \citep{Reames1995, Reames1996, Gopal2006}.
Although the CME ToA has been studied extensively, ToA errors are still high, exceeding 10 hours in many cases \citep{Vourlidas2017, Braga2020}.

Most CME ToA calculations do not take into account deformation. They use CME geometric models that expand self-similarly from the corona to the interplanetary medium \citep{Colaninno2013, Moestl2014, Rollett2016,Moestl2017, Wood2017, Mostl2018}. If we applied these models to the CME studied here, the dimple position would be overestimated. 

Here we estimate the ToA error caused by the deformation. We define deformation as the difference between the CME radial position at the lowest point in the dimple and the corresponding CME reconstructed without deformation at the same latitude. Deformation corresponds to the difference between the squares and crosses in the middle panel of Figure \ref{fig:position_reconst_kinematics}, both calculated in Section \ref{sec:kinematics}. The deformation we get in WISPR-O observations ranges from $0.03\pm 0.01$ au in the first frame to $0.11\pm0.09$ au in the last one. 

\begin{figure}
    \centering
    \includegraphics[width=10cm]{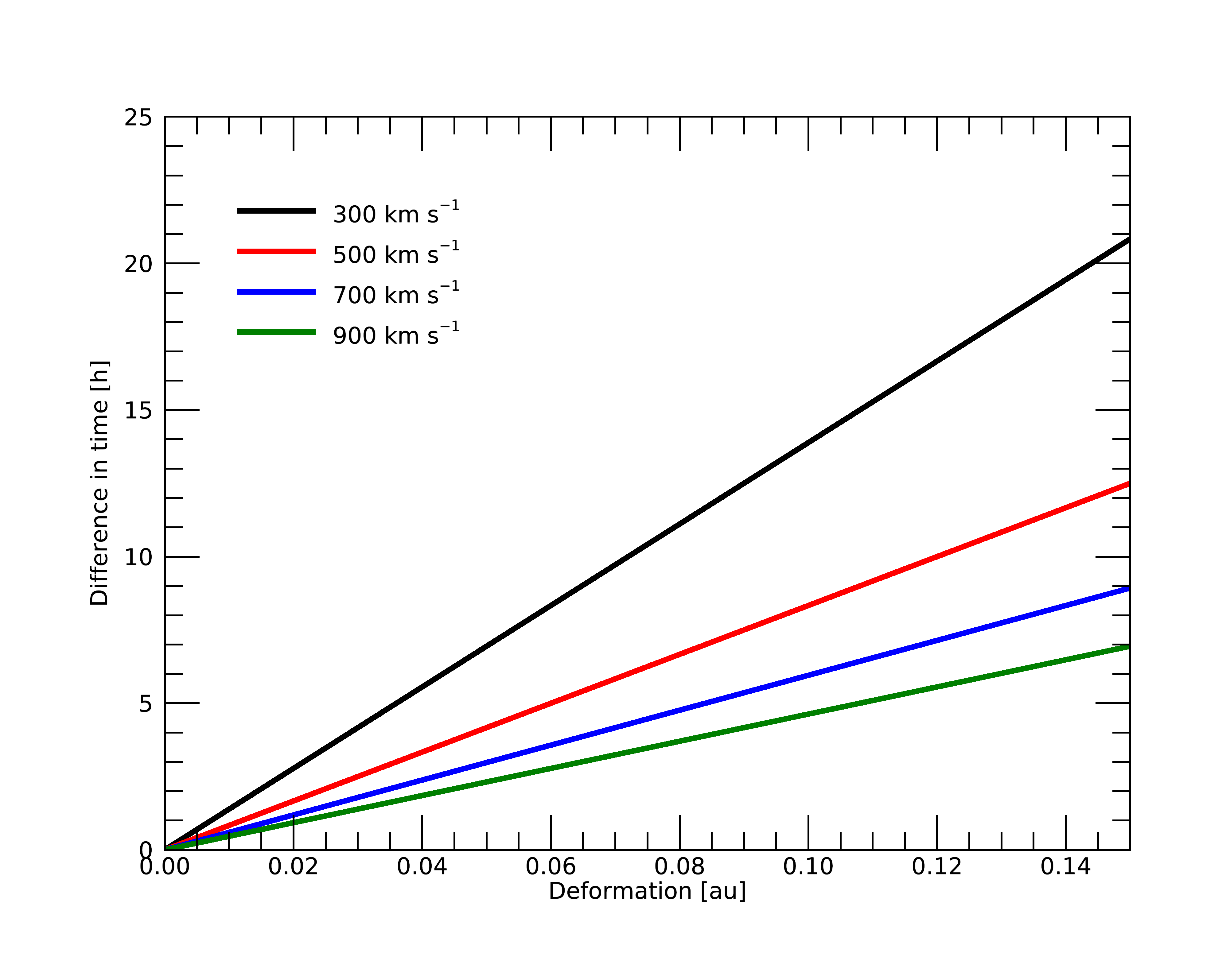}
    \caption{Estimation of the CME Time-of-Arrival (ToA) error that deformation causes. We define deformation as the differences between the actual CME front position in the lowest point in its dimple  ($r_2[t]$, green squares in Figure \ref{fig:position_reconst_kinematics}) with the front position without any deformation (green crosses in Figure \ref{fig:position_reconst_kinematics}). The deformation increases over time and is higher for slow events (black line).}
    \label{fig:deformation_amount}
\end{figure}

We calculate the ToA error caused by the deformation by dividing the deformation with the CME speed. We assume constant speed here as a first-order approximation. The results we get are shown in Figure \ref{fig:deformation_amount}. This estimation considers the case when a hypothetical observer in the central point of the CME dimple. Notice that the error is smaller if the observer is not located at the lowest point in the dimple. 

In the first WISPR-O observation, the CME front is around $\sim0.2\ au$, the deformation is $0.03\pm 0.01\ au$ (Figure  \ref{fig:position_reconst_kinematics}), and we get a 4 hours of ToA error as the current CME speed is $\sim300\ km\ s^{-1}$ (Figure \ref{fig:deformation_amount}). This error is 16 hours for a target close to $0.5\ au$, corresponding to the last WISPR-O observation.

We also calculate the deformation percentage by diving the deformation with the CME leading edge position calculated without deformation. We get $17\pm8\%$ in the first frame and $26\pm22\%$ in the last one. This percentage suggests that the deformation is increasing in WISPR-O FOV, which agrees with the CME shape as seen visually  (Figure \ref{fig:video_wispr_o}). This result  suggests that the CME may continue to deform further from the Sun.

Note that If only coronagraph observations from SOHO and STEREO (left and central columns in Figure \ref{fig:reconstruction3}) were available, the deformation would have probably gone unnoticed. SECCHI/COR2 does not observe the front after $\sim~$03:30, when the deformation begins. Although the CME is still within LASCO/C3 fov when deformation begins, it is not detected clearly. Thus, if we had only SOHO and STEREO observations, we would have the CME ToA errors produced by the deformation.

The lack of deformation in LASCO observations could be explained assuming that it does not extend to the entire CME longitudinal range. The 3 points we tracked are all in the CME portion closer to PSP (Section \ref{sec:kinematics}), which is in the backside from LASCO perspective. If the CME portion close to LASCO is not deformed, the deformation is unlikely to be observed by LASCO. This may happen due to line-of-sight superposition. The instrument observes a superposition of all CME portions along the line-of-sight on a wide range of longitudes as these structures are optically thin. When a fraction of these longitudes does not have deformation, the CME front in the observing instrument might resemble the front of a CME without deformation. Conversely, the background solar wind models suggest that the deformation would take place also in the portion closer to LASCO (right side of the CME in Figure \ref{fig:speed_map}) as the differences in speed and density are clear there.

\section{Summary and conclusions}
\label{sec:summary}

We report an analysis of the first observations of CME deformation by WISPR/PSP. WISPR observes the CME for $\sim 44$ hours (2021-01-20 21:00 to 2021-01-22 17:00). We use forward modeling to reconstruct its morphology over time and to show that the CME is deforming. We also calculate the kinematics of multiple points in the front to evidence the amount of deformation in the central dimple. The results from this analysis are summarized as follows:

\begin{itemize}

\item The CME is slow ($\sim 300\ km\ s^{-1}$) and directed away from the Earth. It is not intercepted by any spacecraft (PSP, SO, STEREO-A, nor L1 observatories). 
\item Our CME reconstruction suggests that the CME deformations starts after 2021-01-21 03:30 when the CME is at 0.1 au ($21\pm4\ R_{\odot}$). Before that time, the CME has a circular profile in the WISPR FOV that can be easily modeled using the graduated cylindrical shell model. The reconstruction also agrees well with SECCHI and LASCO observations. 
\item The CME deformation continues at least until the last available observation available (2021-01-22 around 17:00) where the CME leading edge is at $\sim0.4\ au$. 
\item The deformation seems to be increasing while the CME is between $0.1\ au$ and $0.4\ au$. 
\item The deformation is most likely caused by differences in the background solar wind speed profile. The northern and southern portions of the CME have higher solar wind speeds (and lower density) on MHD coronal models when compared to the CME portion at lower latitudes. 
\item While the CME moves within the likely slow-mode speed range, we cannot determine if the dimple is a slow-mode shock since we lack in-situ measurements in that region. 
\item The dimple is unlikely the result of CME coherence loss. The event  does not meet the criteria for loss of coherence proposed by \cite{Owens2017}.
\item The deformation can result in a CME Time-of-Arrival (ToA) error of 16 hours at $\sim0.5\ au$ for an observer in the CME dimple if we use a model without deformation.
\item The deformation is not observed in the 1 au coronagraphs. The deformation occurs beyond the SECCHI/COR2 FOV and it is not detected in the  LASCO/C3 FOV. The lack of signatures may be due to limited longitudinal extend of the deformation and/or line-of-sight superposition. In either case, the (lack of) observations indicates that CME front deformation may go unnoticed in single-viewpoint observations with important implication for ToA estimates. 

\end{itemize}
As the deformation is evident only in WISPR observations, this raises the question: how many events studied in the past with coronagraph observations may have undetected deformation? This point needs further investigation as it might impact our ability to forecast the CME Time-of-Arrival.
The results from this article can explain why deformation is less common in the coronagraph field-of-view, but more common in interplanetary counterparts of CMEs. 
Another point that needs investigation is understanding what conditions are necessary for deformation to become evident, and at which solar distance they are met.

\begin{acknowledgments}
C.R.B. acknowledges the support from the NASA STEREO/SECCHI (NNG17PP27I) program. A.V. and G.S. are supported by NASA 80NSSC22K0970 and WISPR Phase-E funding. Parker Solar Probe was designed, built, and is now operated by the Johns Hopkins Applied Physics Laboratory as part of NASA’s Living with a Star (LWS) program (contract NNN06AA01C). Support from the LWS management and technical team has played a critical role in the success of the Parker Solar Probe mission. The Wide-Field Imager for Parker Solar Probe (WISPR) instrument was designed, built, and is now operated by the US Naval Research Laboratory in collaboration with Johns Hopkins University/Applied Physics Laboratory, California Institute of Technology/Jet Propulsion Laboratory, University of Gottingen, Germany, Centre Spatiale de Liege, Belgium and University of Toulouse/Research Institute in Astrophysics and Planetology. WISPR data are available for download at  \url{https://wispr.nrl.navy.mil/}. The Sun Earth Connection Coronal and Heliospheric Investigation (SECCHI) was produced by an international consortium of the Naval Research Laboratory (USA), Lockheed Martin Solar and Astrophysics Lab (USA), NASA Goddard Space Flight Center (USA), Rutherford Appleton Laboratory (UK), University of Birmingham (UK), Max Planck Institute for Solar System Research (Germany), Centre Spatiale de Liége (Belgium), Institut d’Optique Theorique et Appliquée (France), and Institut d’Astrophysique Spatiale (France). STEREO/SECCHI data are available for download at https://secchi.nrl.navy.mil/. MAS model output is available from the Predictive Science Inc. website: \url{https://www.predsci.com/hmi/data_access.php} 

\end{acknowledgments}

\appendix
\label{sec:appendix}

Here we briefly explain the processing technique we use to derive WISPR images displayed in this paper (Figures \ref{fig:video_wispr_i},  \ref{fig:video_wispr_o}, \ref{fig:reconstruction23}, \ref{fig:reconstruction3}, \ref{fig:cme_distorted}, and \ref{fig:wispr_o_3_points_location}), and for STEREO/COR2 images in Figures \ref{fig:reconstruction23} and \ref{fig:reconstruction3}. Briefly, this technique enhances the CME visualization by determining and removing the background of each image, which features features other than the CME, mainly the F-corona formed by the dust and some components of the K-corona that appear stationary in the FOV of the instrument. 

To compute the background scene of WISPR images, we exploit the time domain. This approach is carried out independently at each pixel location, by estimating the running 5-percentile basal brightness level. This level is strongly influenced by the F-corona component, which exhibits a relatively large variation in time due to the varying heliocentric distance of the observer. Therefore, detrending of the time series is needed. For the particular case of the WISPR images used in this paper, the detrending was performed by normalizing the time series with a smoothed version of the median filtered time series. The running median filter was applied using a window length equivalent to the time elapsed by seven images. The smoothing is obtaining by convolving the median-filtered time series with a uniform, 1D kernel of size equivalent to the time elapsed during nine images. The length of the running window and kernel size affects the degree of detail in the resulting product, akin to a high pass filter. For details the reader is referred to Howard et al. (in preparation). A sigma-filter as implemented in IDL ({\tt\string sigma$\_$filter.pro}) was applied to the background corrected frames displayed in Figures \ref{fig:video_wispr_i}, \ref{fig:video_wispr_o}, and \ref{fig:wispr_o_3_points_location} to reduce the effect of point-like sources like, e.g., the star field. WISPR images in Figures \ref{fig:reconstruction23}, \ref{fig:reconstruction3}, and \ref{fig:cme_distorted} are processed in a similar way, but without the sigma filter. We also use different level adjustments in these images.  

We also use a especial imaging processing technique for STEREO/COR2 to enhance the CME visualization in Figures \ref{fig:reconstruction23} and \ref{fig:reconstruction3}. Unlike the images obtained with PSP/WISPR, STEREO/COR2 images are obtained from a relatively constant distance. In spite of this, we applied a similar procedure to remove the background while trying to preserve pseudo-stationary structures like streamers (unlike WISPR case where we removed them). With that aim in mind, the detrending was carried out by robust-fitting the time series with a straight line. 

\bibliography{sample631}{}
\bibliographystyle{aasjournal}



\end{document}